%% file: chicago_elsevier.tex
\documentclass[preprint,12pt,authoryear]{elsarticle}

\usepackage{amssymb}
\usepackage{makecell}
\usepackage{xcolor,tabularx}
\usepackage{booktabs,caption,subcaption,mathtools,multirow,graphicx}

\usepackage{amsmath,amsthm,enumitem}

\usepackage{soul}

\usepackage{algorithm}
\usepackage{algpseudocode}

\usepackage[section]{placeins}
\usepackage{url}

\input{common_preamble.tex}

\renewcommand{\figscale}{1.25}

\renewcommand{\tabfit}{\setlength{\tabcolsep}{3pt}\small}

\journal{Transportation Research Part D: Transport and Environment}

\begin{document}

\begin{frontmatter}

\title{\papertitle}

\author[anl]{Gopindra Sivakumar Nair\corref{cor1}}
\cortext[cor1]{Corresponding author}
\ead{gnair@anl.gov}
\author[lbnl]{Yilin Jiang}
\author[usim]{Samuel Maurer}
\author[anl]{James Cook}
\author[anl]{Nazmul Arefin Khan}
\author[anl]{Joshua A. Auld}
\author[lbnl]{Tianzhen Hong}
\author[usim]{Arezoo Besharati}
\author[usim]{Paul Waddell}

\affiliation[anl]{organization={Argonne National Laboratory}, city={Lemont}, state={IL}, country={USA}}
\affiliation[lbnl]{organization={Lawrence Berkeley National Laboratory}, city={Berkeley}, state={CA}, country={USA}}
\affiliation[usim]{organization={UrbanSim Inc.}, city={Berkeley}, state={CA}, country={USA}}

\begin{abstract}
\input{abstract_elsevier.tex}
\end{abstract}


\end{frontmatter}

\input{body.tex}

\input{backmatter.tex}

\bibliographystyle{elsarticle-harv}
\bibliography{chicago_cosim}

\end{document}

%% file: common_preamble.tex
\graphicspath{{paper_figures/}{figures/}}

\newcommand{\papertitle}{A Co-Simulation Platform Coupling Land Use, Transportation, and Building Energy: Development and Case Study}

\defcitealias{nhts2017}{FHWA}

\providecommand{\figscale}{1}
\newcommand{\figw}[1]{\fpeval{min(1, #1 * \figscale)}\linewidth}

\providecommand{\tabfit}{}

\newif\ifdraft \drafttrue

%% file: abstract_elsevier.tex
Land use, transportation, and building energy shape one another, yet urban-scale studies typically model each sector in isolation. This work develops a co-simulation platform that runs the three sectors in an integrated workflow and produces internally consistent forecasts, and demonstrates it by tracing how two transportation scenarios propagate into land use and building energy. The platform couples the UrbanSim land-use model, the POLARIS agent-based transportation model, and the CityBES urban building energy model, with POLARIS travel skims driving land use and POLARIS agent activities driving dynamic building occupancy. Forecasts were produced through 2045 for the Chicago metropolitan area under a business-as-usual case, a high-telecommuting scenario, and a mileage-based user fee scenario. Both policies produced expected-direction responses that emerged from the model feedbacks rather than being imposed. Telecommuting decentralized activity toward outlying areas and cut 2045 vehicle miles traveled by 12.5\%, whereas the mileage fee recentralized activity toward the urban core and cut it by 2.9\%. Comparing coupled runs against uncoupled runs that hold land use fixed showed that the land-use feedback contributes over one percentage point to the county-level travel effect in several counties. This is large relative to the policy effect itself, since the mileage fee's county effects are only about three percent, so a transportation-only study would materially misstate the sub-regional impact. Both policies raised citywide building energy by about 1\%. This is the first platform to integrate land use, transportation, and building energy simultaneously, replacing predefined occupancy schedules and static building stocks with endogenous agent-based occupancy and a forecast-driven building stock. It lets planners evaluate transportation and pricing policies for their joint land use, travel, and energy consequences, and its component models rely on nationally available data, so it is transferable given local building-stock and calibration data.

%% file: body.tex
\section{Introduction}\label{sec:intro}
Buildings and transportation together account for $70\%$ of energy use in U.S. cities \cite{EIA_AEO}, and both are strongly influenced by the underlying land-use pattern of where people live, where they work, and how they travel between the two. Yet in most urban-scale studies these three sectors are modeled separately, often by different teams using different tools, so the outputs of one model only loosely inform the inputs of the next. Population and employment distributions are typically exogenous in transportation forecasting, building energy models assume standardized occupancy curves, and even land-use models that ingest travel skims are rarely run alongside transportation models in an integrated fashion. This separation stems less from conceptual disagreement about the interactions than from a disconnect between the independent software platforms and developer communities underlying each field \cite{hong_10Q_UBEM,miller_2018}. Representing all three domains within a single platform preserves the fidelity of each component while enabling the feedbacks that sequential, one-way modeling cannot capture, such as accessibility shaping location choice and occupancy shaping building energy demand, so that the cascading effects of cross-sectoral policies can be traced across all three domains at once. Efforts to overcome this separation have followed two broad directions. Data-driven frameworks predict building and transportation energy jointly from shared urban attributes \cite{abbasabadi_integrated_2019}, while bottom-up approaches co-simulate the physical sector models directly \cite{berres_mobility_2019}. Reviews of multi-domain urban energy modeling identify the binding difficulty as bridging data across sectors while still respecting the physics each sector must obey \cite{bishop_review_2024}.

The remainder of this section reviews prior work along the three pairwise couplings between land use, transportation, and building energy, and then states the contributions of this study.

\subsection{Land Use and Transportation Integration}\label{subsec:landuse_transportation}

Of the three pairwise couplings, land use and transportation has the most established modeling tradition. The two are linked through a feedback in which the transportation network determines accessibility, shaping where households and firms locate, while the resulting distribution of activity generates the travel demand the network must accommodate \citep{acheampong_review_2015}. Operational systems have represented this cycle since the 1960s, and UrbanSim was developed to serve it within regional transportation planning \citep{waddell_urbansim_2002, waddell_integrated_2011}, including policies such as transit-oriented development \citep{keirstead_sivakumar_2012}. Such systems have been used to trace how network changes redistribute households and firms and feed back into travel demand \citep{zhou_kockelman_lemp_2009, moeckel_trends_2018}. UrbanSim and POLARIS in particular have been run in a closed loop under the U.S. Department of Energy SMART Mobility program \citep{rousseau_smart_2020}, and \citet{auld_et_al_2026} used that framework to show that explicitly modeling household and workplace location choice changes long-term estimates of vehicle miles traveled (VMT) under vehicle automation. However, these systems remain confined to the transportation sector and do not propagate the built-environment forecasts of the land-use model into a physics-based energy model.

\subsection{Transportation and Building Energy Integration}\label{subsec:transportation_energy}

Building occupancy is a key input to building energy loads \cite{alfalah_et_al_2023}, yet occupant presence is commonly modeled with predefined a priori schedules \cite{brian_et_al_2017}. Applying the same schedule across all buildings of a type can introduce large errors. \citet{duarte_et_al_2013}, for example, found measured peak occupancy for private offices up to 46\% below the ASHRAE reference profile. Because occupancy strongly influences lighting, plug loads, ventilation, and heating and cooling demand, such mismatch degrades energy estimates.

More realistic occupancy can be obtained by integrating a physics-based building energy model with an agent-based travel model, though existing efforts remain partial. \citet{berres_mobility_2019} coupled a synthetic population and National Household Travel Survey commute patterns (\citetalias{nhts2017}, \citeyear{nhts2017}) to TRANSIMS traffic assignment, but were limited to commute trips with departure times insensitive to congestion. \citet{mosteiro_et_al_2020} assigned activity schedules with a MATSim-inspired tool in which movements between activities did not account for congestion or mode availability, and \cite{hou_et_al_2026} modeled in-building activity in detail for a college campus but with a spatially constrained travel model whose disruptions were exogenous. Others infer occupancy from mobility data without modeling activity and travel demand, distributing origin--destination trips to buildings \citep{pan_roads_2024} or deriving occupancy from multi-modal flows \citep{abbasabadi_integrated_2019}. These approaches capture spatial and temporal variation absent in fixed schedules, but assign occupancy by proximity or aggregate flow and treat the building stock as static, so they do not represent the dynamic interaction with land use.

\subsection{Building Energy and Land Use Integration}\label{subsec:building-landuse}
A building energy model needs a building stock to simulate, which urban building energy models assemble from GIS footprints, LiDAR returns, and tax-assessor records of the buildings already in place, joined to an archetype or prototype library \citep{chen_hong_UBEM_dataset, zhou_ubem_2025, ang_concept_2020}. CityBES, AutoBEM, and GeoBEM build the present-day stock in this way \citep{hong_citybes_2016, bass_potential_2021, zhang_geobem_2025}. Urban building energy modeling is often framed as bridging building science and urban planning \citep{hong_10Q_UBEM,zhou_ubem_2025}, yet land use mostly feeds it in one direction and is used to describe the existing stock rather than to generate a prospective one. Zoning has been used to classify an existing office stock into typologies for an energy baseline \citep{alves_methodology_2017}, and land-use mix and density have been related to the energy of an existing regional stock under renovation and demographic scenarios \citep{nishimwe_energy_2023}. In both cases the stock modeled is the one already developed, and the new buildings a land-use forecast implies are never actually generated.

The reverse direction is similarly underdeveloped. Land-use and integrated urban models have long represented the real-estate development process. UrbanSim, for instance, simulates developer decisions on what to build, where, and at what density \cite{waddell_urbansim_2002}, and integrated urban modeling has been advocated precisely to capture such cross-sector dynamics \cite{miller_2018}. Yet, the buildings these models produce are described in the aggregate terms of land-use analysis, as dwelling units, floorspace, or jobs accommodated, rather than as geometrically and thermally resolved structures a physics-based energy model requires. Bridging this representational gap, from forecast dwelling units and jobs to discrete buildings with a defined type, vintage, footprint, and location, is the task of the building generation module developed in this study.

\subsection{Contribution}
In this paper, we develop a comprehensive workflow that integrates the UrbanSim land-use model \cite{waddell_urbansim_2002}, the POLARIS agent-based transportation model \cite{auld_polaris_2016}, and the CityBES building energy model \cite{hong_citybes_2016, chen_auto_UBEM}, all anchored on the same geospatial data layer \cite{bureau_2020_nodate}. In this integrated framework, land-use evolution in UrbanSim is driven by generalized skim outputs from POLARIS, while the spatiotemporal movement of individuals in POLARIS is translated into dynamic building occupancy patterns for CityBES. While previous studies have separately coupled transportation with land use or transportation with building energy, this paper presents the first attempt at integrating models across all three sectors simultaneously. To achieve this, we bridged the geographic resolution gap between UrbanSim, a block-level macro model, and POLARIS, which requires person- and household-level inputs. We enhanced the native POLARIS population synthesizer to update the agent population based on future-year and block-level marginals generated by UrbanSim. Furthermore, unlike prior building-energy integration efforts that utilized only specific aspects of transportation modeling, such as traffic assignment, or relied on simplistic models where even the effects of network conditions were treated as exogenous inputs, our approach leverages the full-fledged agent-based activity generation and routing capabilities of POLARIS to endogenously capture the impacts of congestion, mode choice, and complex travel behavior on building energy use. The co-simulation workflow was used to generate forecasts through the year 2045 for the Chicago study area under three scenarios. The first is a BASE scenario representing business-as-usual changes between 2025 and 2045. The second is a telecommuting scenario (TELE) in which the daily telecommuting rate in 2045 is 29\%. The third is a mileage-based user fee (MBUF) scenario, in which vehicles are charged an additional fee of \$0.06 per mile traveled. 

\section{Methodology}\label{sec:methodology}

\subsection{Co-simulation platform overview}\label{subsec:platform}
\begin{figure}[!ht]
  \centering
  \includegraphics[width=\figw{0.82}]{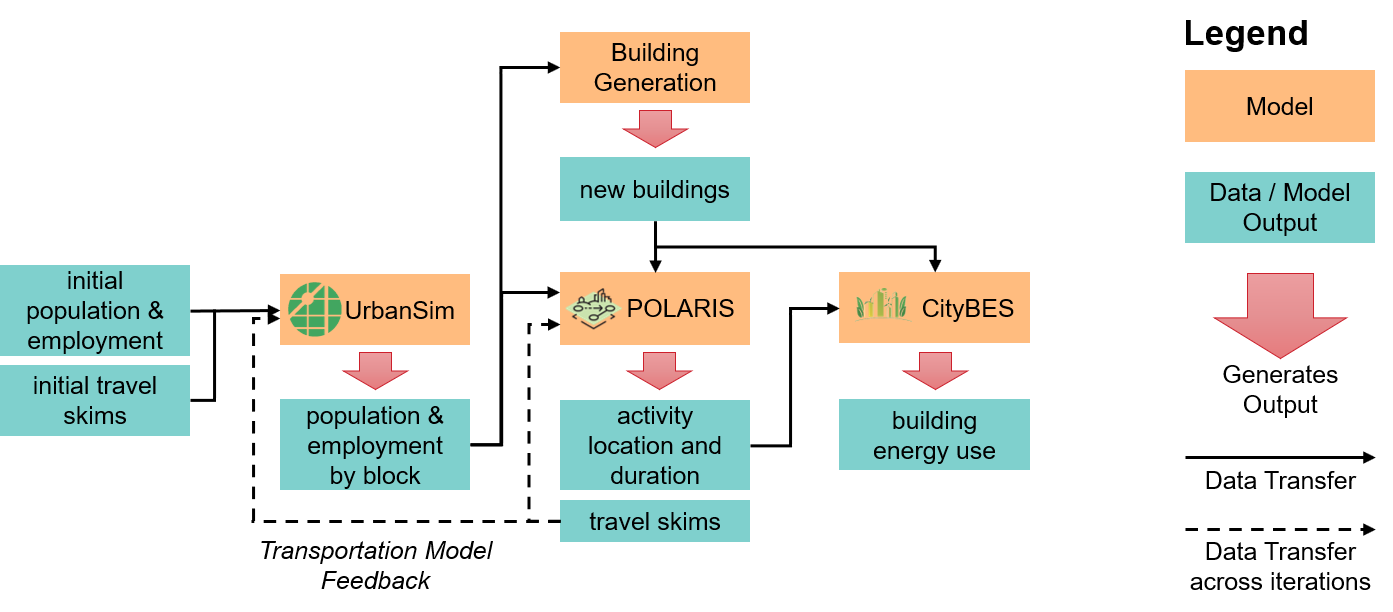}
  \caption{Co-simulation platform structure}
  \label{fig:overview}
\end{figure}

Figure \ref{fig:overview} illustrates the structure of the co-simulation platform, where UrbanSim, POLARIS, and CityBES run iteratively in sequence. UrbanSim updates population and employment annually, whereas POLARIS and CityBES are executed at five-year intervals. At the start of the process, the base year population and employment are obtained from UrbanSim and passed to POLARIS, which generates skims along with building occupancies derived from activity locations and durations. The converged skims from POLARIS feed back into UrbanSim to evolve the land use over the next five years, and they also warm start the iterative convergence process in POLARIS for the next forecast year. The occupancies are sent to CityBES to generate building energy use patterns. In forecast years, the updated land use forecasts from UrbanSim are first converted to new buildings that are added to the Chicago building stock used by CityBES and to the set of locations POLARIS can assign activities to. The population synthesizer in POLARIS then synthesizes additional persons and adds them to the previous year's population to meet the new block-level population totals from UrbanSim for the current year, removing persons instead where those totals shrink.

\subsection{Component modules}\label{subsec:components}

\subsubsection{Land use: UrbanSim}\label{subsec:urbansim}
UrbanSim is a simulation model of urban development that forecasts the evolution of population, employment, and the built environment in response to transportation conditions, market dynamics, and policy \cite{waddell_urbansim_2002}. The deployment in this study is a modification of UrbanSim's model for the Chicago Metropolitan Agency for Planning (CMAP). It represents the study area as 2020 Census blocks and simulates the location choices of households and jobs at that resolution. The Chicago region model covers 143{,}999 blocks.

Within this platform, UrbanSim runs as a cloud service: its base data, calibrated parameters, and annual outputs are maintained in dedicated cloud storage. Key inputs and outputs are referenced directly in API calls, so the orchestrator advances the land-use simulation one year at a time without a local installation. In each annual step, UrbanSim carries forward the prior-year distribution of households and jobs, applies regional control totals, and evaluates accessibility from the most recent POLARIS travel skims to produce block-level forecasts of households, jobs, and residential units for the study area.

Each simulated year proceeds in three stages. First, regional control totals (growth expectations) determine the total count of households, persons, and employment by two-digit NAICS sector. Second, transition and relocation models determine which households and jobs enter the region, exit, or become candidates for relocation. Third, location choice models allocate the moving and newly arriving agents to blocks, as well as adding or removing housing stock. Job and housing growth is subject to block-level zoning capacity derived from local land-use regulation.

The behavioral models are logit models over blocks, estimated separately by market segment, with households segmented by life stage, income group, and tenure and employment by two-digit NAICS sector. Utility functions combine local attributes such as price or rent, competing units or jobs, developable area, and the surrounding income composition with regional accessibility measures computed from the POLARIS travel skims described in Section~\ref{subsec:urbansim-polaris}, and prices and rents update endogenously through demand-to-supply adjustment. Accessibility enters the location-choice and development models as a cumulative-opportunity count of jobs reachable within an impedance threshold, for both automobile and transit. Because the travel skims are refreshed after every five-year POLARIS run, changes in network conditions propagate into these accessibility terms and alter subsequent location choices, which is the mechanism by which the transportation scenarios studied here produce diverging land-use trajectories.

For each simulation year, UrbanSim writes block-level tables of households, cross-classified by life stage, income group, and tenure, and jobs by two-digit NAICS sector. These tables serve simultaneously as the marginals for the differential population synthesis of Section~\ref{subsec:urbansim-polaris} and as the inputs to the building generation module of Section~\ref{subsec:building-generation}.

\subsubsection{Transportation: POLARIS}\label{subsec:polaris}

POLARIS is an agent-based, activity-based transportation modeling framework for the high-performance simulation of large metropolitan regions \cite{auld_polaris_2016}. Given marginal totals it synthesizes a population, schedules and plans each agent's activities under spatiotemporal constraints, and progresses through the day materializing those activities. The resulting trips are routed with a time-dependent dynamic traffic assignment router \cite{verbas2018time}, and a multi-class traffic flow model determines dynamic link travel times \cite{de2024polaris}, so congestion feeds back into subsequent choices and endogenously captures the effects of network conditions, mode availability, and travel time on behavior. Because activities are resolved for each agent with explicit locations, start times, and durations, POLARIS produces the detailed spatiotemporal record of where people are throughout the day that the co-simulation uses to construct dynamic building occupancy (Section~\ref{subsec:occupancy}). Each year is itself solved iteratively. Starting from the prior year's skims, agents plan and execute activities, updated skims are computed from the resulting network conditions, and agents re-plan and reroute until convergence. The prior year's skims warm-start the iterations, so fewer are needed.

\subsubsection{Building energy: CityBES}\label{subsec:building}
CityBES is an urban building energy modeling (UBEM) platform that represents a city's stock with 3D CityGML models, simulates the buildings in EnergyPlus through an OpenStudio-based software development kit, and returns hourly per-building energy \cite{hong_citybes_2016}. It has supported city-scale studies of peak-load forecasting, heat-vulnerability assessment, and passive-cooling design \cite{zhang_calibration_2026, xu_heat_2025, sun_passive_2021}. We run it through a CityBES API developed for the platform. Unlike the CityBES web application, which models every physical building explicitly, the API simulates each building prototype once and scales the result to the count of buildings of that type, keeping a repeated, city-scale co-simulation tractable. The module supports twenty-one building types across the residential, commercial, institutional, and industrial sectors, spanning single-family and multifamily homes, small to large offices, quick- and full-service restaurants, supermarket, strip-mall, and standalone retail, hotels, primary schools, secondary schools, and colleges, nursing homes, hospitals, and outpatient facilities, courthouses, and heavy manufacturing, light manufacturing, and warehouses. Each type is resolved further by vintage, using eight ASHRAE~90.1 bins from pre-1980 to 2019, differentiated by the transportation-informed occupancy of Section~\ref{subsec:occupancy}, and simulated against location-specific weather.



\subsection{Integration}\label{subsec:integration}

\subsubsection{UrbanSim and POLARIS}\label{subsec:urbansim-polaris}

Block-level marginals from UrbanSim are used as inputs to the POLARIS population synthesizer. For the base year, a direct population synthesis is performed using these marginals. In subsequent simulation years, once the forecast-year marginals are obtained, they are compared with the marginal totals of the existing POLARIS population. Any population difference is synthesized and appended to the existing population. In some cases, the marginals for a block in a forecast year may be lower than those in the previous year. In such cases, households are randomly removed from the block until all marginal totals are nonnegative, after which additional population is synthesized as needed to match the forecast marginals.

POLARIS also represents the long-term decisions that link travel behavior to land use. For each agent, the time spent in the current home and workplace is initialized in the base-year run and then updated each year with hazard-duration models that determine whether a change occurs \cite{bostanara2021comparison}. When a workplace change is indicated, POLARIS applies its workplace choice model to assign jobs, and the same model assigns workplaces to newly synthesized individuals. 

After the POLARIS run is completed, the AM peak-period auto and transit skims are passed back to UrbanSim for use in subsequent forecast years.

\subsubsection{UrbanSim and CityBES}\label{subsec:building-generation}
The interaction between UrbanSim and CityBES is mediated by a building generation module that produces the new construction that building energy simulation requires. UrbanSim reports its forecast as dwelling units and jobs by NAICS sector at the block level, and the module turns these aggregate counts into discrete new buildings, each with a building type, a location, and a floor area, so that CityBES can simulate them. For residential buildings, it compares the change in dwelling units between the baseline year $t_0$ and the simulation year $t_1$, and assigns a building type from the block-level dwelling-unit density, using thresholds drawn from city-scale survey datasets. For commercial buildings, it compares the change in jobs between $t_0$ and $t_1$ and allocates them with a procedure that rests mainly on the employment density of each building type. The residential and commercial allocators, their parameters, and their validation against an independent Chicago building stock are detailed in \cite{jiang_coupling_2026}.

\subsubsection{POLARIS and CityBES}\label{subsec:occupancy}
 
POLARIS informs CityBES by supplying the building-level occupancy that building energy simulation requires. POLARIS simulates each individual's daily schedule, including the location, type, and duration of every activity. A building-to-activity map, built from the Chicago building stock, marks which parcels can host each activity, and POLARIS places activities only on the allowed parcels, producing a dynamic parcel-level occupancy. For example, a parcel with only restaurants admits only work and eat-out activities, an office parcel admits only work, and a supermarket admits work and shopping. The occupancy in a parcel is then distributed among the buildings within it in proportion to each building's floor area.

\section{Case Study}\label{sec:case-study}

\subsection{Study Geography}

The Chicago metropolitan area constituted the study area. Because the data required for calibrating and validating the component models varied in availability, the geographical extent of each model's study area differed. The UrbanSim model covered the seven counties of the CMAP region: Cook, DuPage, Kane, Kendall, Lake, McHenry, and Will. The POLARIS transportation model had the largest extent, covering at least 11 additional counties beyond those modeled by UrbanSim. Because land-use changes in the outer counties were not modeled in UrbanSim, those counties were simulated in less detail in POLARIS, with their population and employment held constant; they were nonetheless retained in the POLARIS simulation to produce realistic traffic entering the core region. The Chicago building-stock data required for building energy analysis was available only within the Chicago city limits, so building energy modeling was performed only within the city.

The UrbanSim forecast is anchored to regional control totals that are held fixed across scenarios. We use population and employment projections from the draft 2026 Regional Transportation Plan prepared by CMAP \cite{cmap2026rtp}. Regional population grows modestly and monotonically over the horizon, from about 8.61 million in 2025 to about 8.82 million in 2045, an increase of roughly 2.5\%, with the growth stalling toward the end of the horizon after 2040. Regional employment follows a different path, rising slightly to a peak of about 4.61 million in 2030 and then declining steadily to about 4.54 million by 2045 as the population ages, a net decline of roughly 1.5\% over the period.

\subsection{Scenarios}

Three scenarios were run with the co-simulation platform: a business-as-usual base case (BASE) and two transportation-policy scenarios, high telecommuting (TELE) and mileage-based user fee (MBUF). All three share a common simulation stem from 2025 through 2030, following the sequence described in Section~\ref{subsec:platform}. To reflect a realistic adoption timeline, scenario-specific conditions are introduced in the travel model only from 2030 onward, at which point the three scenarios branch from the shared stem. From 2030, the scenario-specific travel skims produce diverging land-use patterns in UrbanSim, and the scenario-specific occupancies drive differences in building energy use in CityBES.

\subsubsection{Business as usual (BASE)}

This scenario represents a continuation of current conditions with no policy intervention and serves as the reference against which the two policy scenarios are measured. It is anchored to the regional control totals described above and will henceforth be referred to as BASE.

\subsubsection{High Telecommuting (TELE)}

This scenario represents a sustained shift toward remote work in which the daily telecommuting rate reaches 29\% by 2045 \cite{danalet2021working}, up from a baseline daily telecommute rate of 8\%. The baseline telecommute model in POLARIS was calibrated using the 2018-19 CMAP Household Travel Survey through systematic adjustment of parameters via an industry-based approach that controlled for both employment industry classifications and alternative-specific constants (ASCs) to closely match the calibration target. This high telecommuting scenario will henceforth be referred to as TELE.

\subsubsection{Mileage based user fee (MBUF)}

This scenario mimics the mileage-based tax being considered in several states \cite{shrode2023vmt}. The tax is meant to compensate for the reduced tax revenue from fuel sales as vehicles become more efficient. In this study we used a fee of 6 cents per mile which was applied at the link level in POLARIS similar to toll costs. The mileage fee is modeled based on the distance based fee Austin simulation in \citet{mori2026comparing}. The rate is used as a representative policy benchmark rather than as an estimate of the socially optimal fee. This scenario will henceforth be referred to as MBUF.

\section{Results}\label{sec:results}
The co-simulation platform was run on the Argonne National Laboratory's Laboratory Computing Resource Center (LCRC) compute nodes, each having 512GB RAM and 128 cores. Each node was used to run two simulations in parallel. UrbanSim and CityBES modules were executed through API calls while POLARIS was run locally. The simulations ran in approximately 44 hours per forecast year. This runtime is dominated by running POLARIS to equilibration through 14 iterations (about 43.5 hours), with the remainder spent on land use evolution (0.5 hours). Generating city-wide, building-level results with CityBES takes roughly 4 hours per run. CityBES was executed in parallel, with inputs submitted through API calls so that the co-simulation continued without waiting for its results.

\subsection{Land-use response}\label{subsec:results-landuse}

UrbanSim reallocates population and employment relative to the BASE forecast rather than changing regional totals, so the land-use response to each policy is a spatial pattern rather than a change in how much the region grows. The model has no direct representation of telecommuting or of a mileage fee. Instead, these come through via the POLARIS travel skims and shift the relative attractiveness of different locations.

\begin{figure}[!ht]
  \centering
  \includegraphics[width=\figw{0.75}]{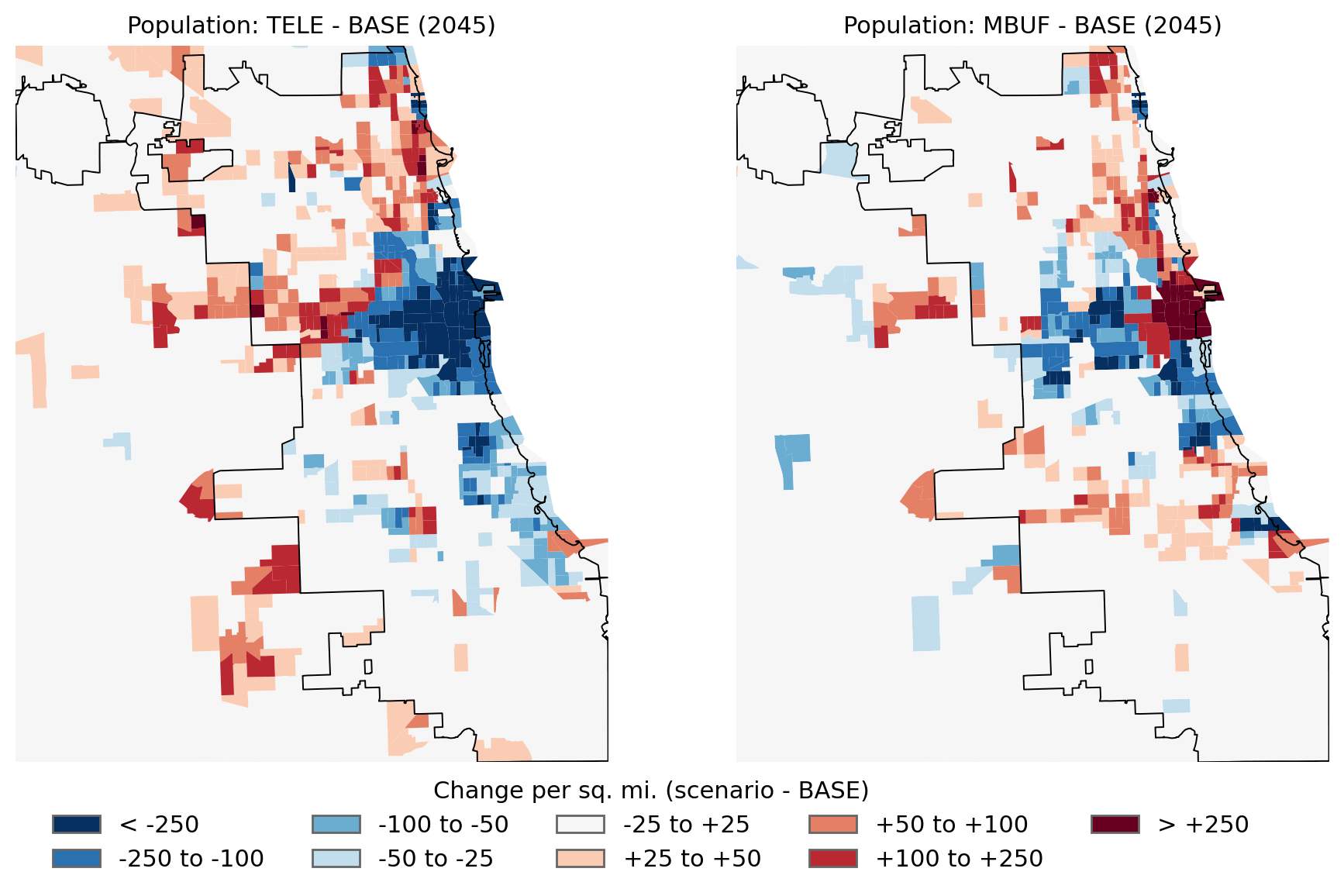}\\[2pt]
  \includegraphics[width=\figw{0.75}]{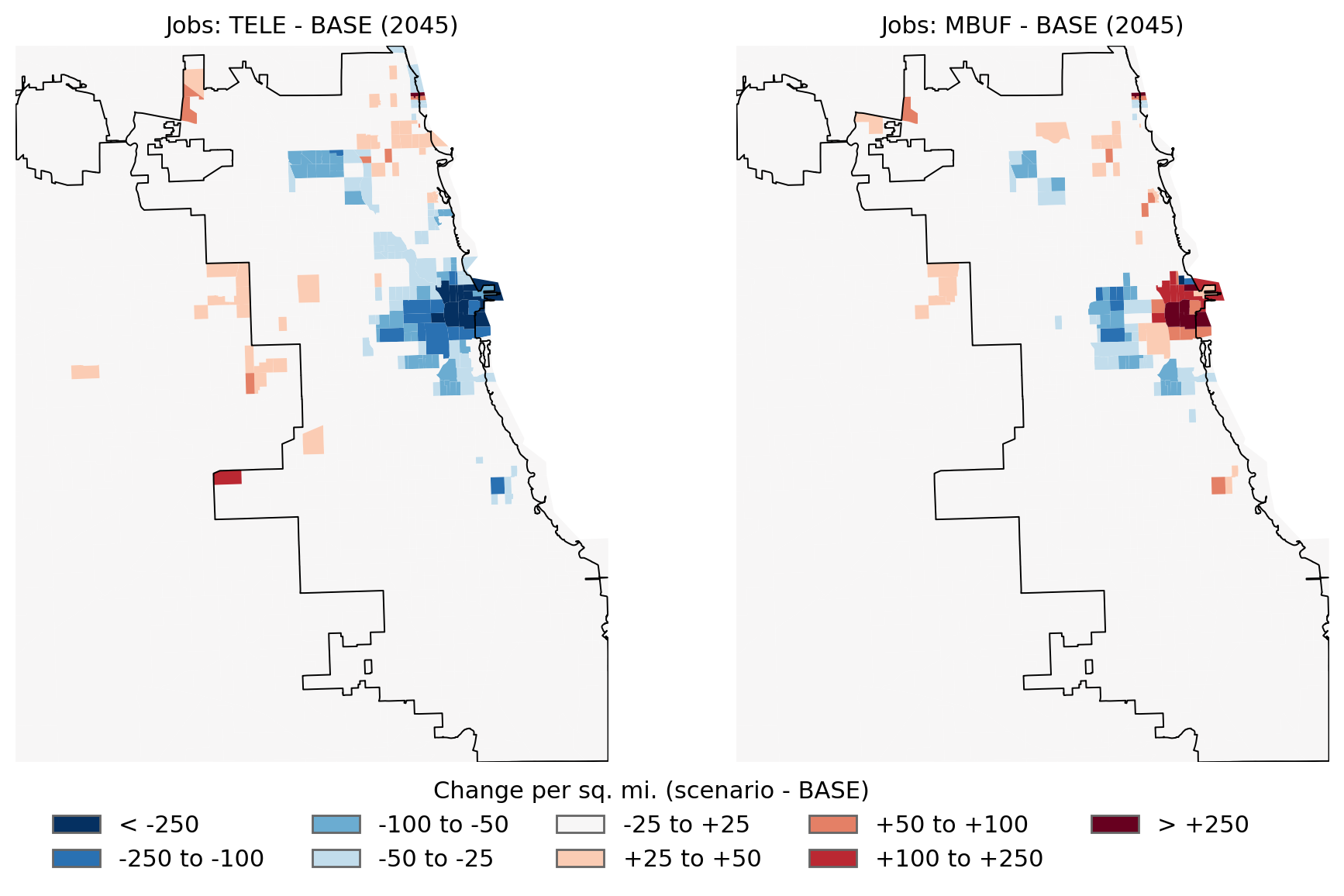}
  \caption{Tract-level change relative to BASE in 2045 for (a, top) population density and (b, bottom) job density, TELE (left) and MBUF (right), shown for the bounding box of the City of Chicago.}
  \label{fig:usim-maps}
\end{figure}

The reallocation is best seen at fine spatial resolution, because county or even city aggregates net out large offsetting movements and make the response look muted. Although the model operates at the census-block level, we aggregate the results to the census tract for presentation. Figure~\ref{fig:usim-maps} maps the 2045 change relative to BASE in population density (top) and job density (bottom) across the City of Chicago.

The two policies move activity in opposite directions. Under TELE the dense urban core thins almost uniformly while outlying tracts gain. Fewer daily trips relieve network congestion and reduce travel times across the region, which disproportionately improves the relative accessibility of outlying locations, so UrbanSim places more growth there. Under MBUF the same core gains at the expense of the periphery, because the per-mile fee raises the generalized cost of the long trips associated with dispersed locations and strengthens the relative advantage of the dense, high-accessibility core.

\subsection{Travel demand response}\label{subsec:results-travel}

\begin{figure}[!ht]
  \centering
  \includegraphics[width=\figw{0.8}]{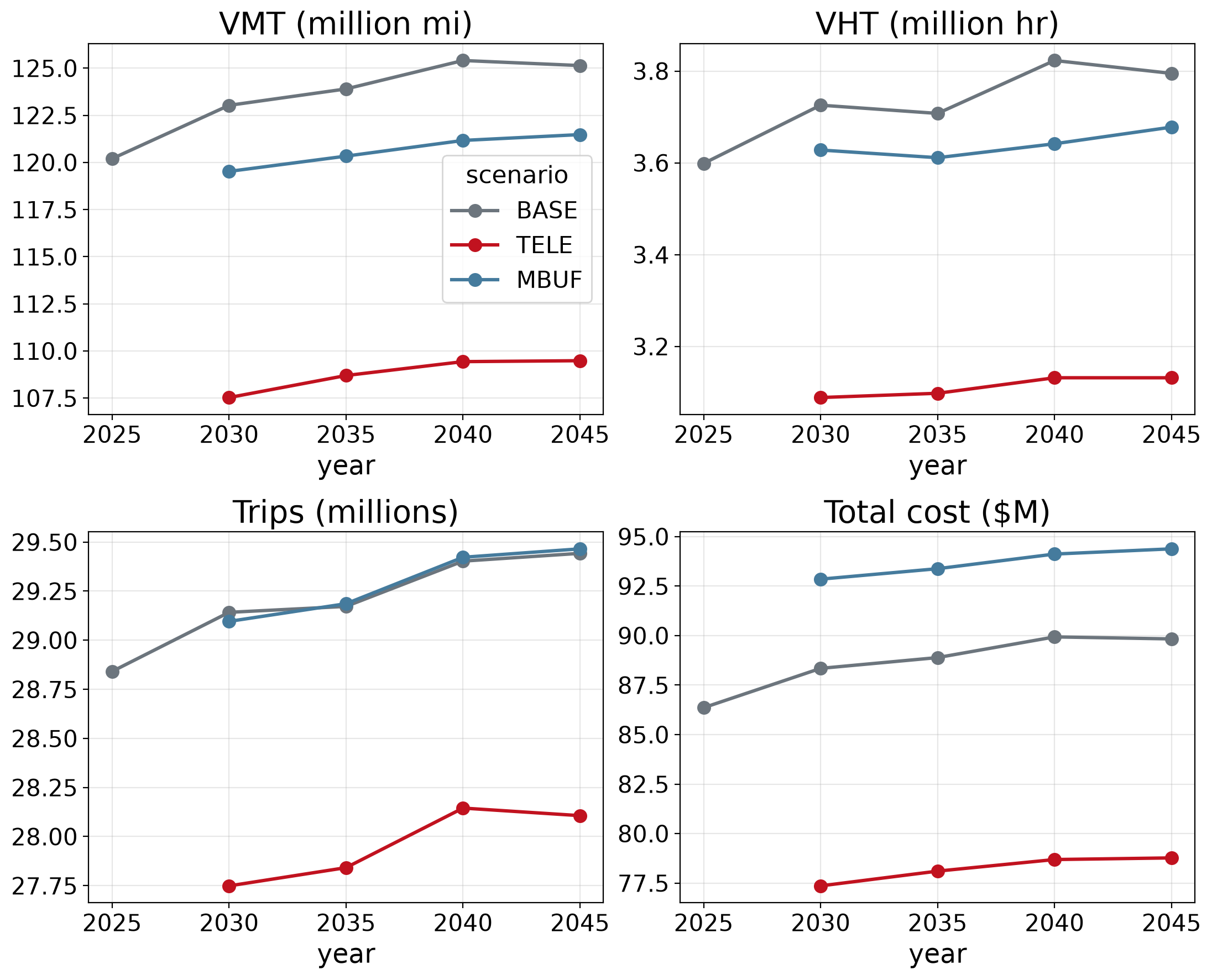}
  \caption{Household travel over the forecast horizon for the coupled scenarios, seven-county core.}
  \label{fig:travel-metrics}
\end{figure}

We summarize household travel over the seven-county simulated core. The metrics included are Vehicle Miles Traveled (VMT), Vehicle Hours Traveled (VHT), the number of trips, and total cost. Total cost sums fares, tolls, and an auto operating cost applied to Auto-Driver miles at the \citet{irs_notice_2024_08} standard mileage rate of \$0.67 per mile, which covers fuel, maintenance, depreciation, and insurance. The mileage fee under MBUF enters through the toll term. Figure~\ref{fig:travel-metrics} traces the regional totals over the forecast horizon. The two policies move travel in the same direction but by very different magnitudes. By 2045 TELE lowers VMT by about 12.5 percent and VHT by about 17.5 percent relative to BASE, while MBUF lowers VMT by about 2.9 percent and VHT by about 3.1 percent. In both scenarios VHT falls proportionally more than VMT, so average network speed rises.

The trip count separates the two policies along a different margin. TELE removes trips outright, lowering the number of trips by about 4.5 percent by 2045, yet its VMT falls almost three times as steeply because the trips that remain are on average shorter. MBUF leaves the number of trips essentially unchanged, within a tenth of a percent of BASE, while still trimming VMT by about 2.9 percent. The mileage fee therefore does not deter households from traveling but shortens and reroutes the driving they do, whereas telecommuting acts on the extensive margin by eliminating commute trips.

The BASE trajectory itself is not monotonic. VMT and VHT rise to about 2040 and then edge down by 2045. This tracks the UrbanSim control totals rather than any travel-side artifact. Across the forecast years, regional VHT is almost perfectly correlated with the core-county population held in the control totals (Pearson $r = 0.97$, and $r = 0.98$ against population plus employment). The late-horizon decline coincides with population growth stalling after 2040 while employment declines steadily from its 2030 peak, which reduces commute travel even as population plateaus. The land-use and travel results are therefore internally consistent.

The cost response distinguishes the two policies most sharply. Decomposing 2045 travel cost into fares, tolls, and auto operating cost shows that the operating cost of driving dominates the total in every scenario. TELE lowers total cost roughly in proportion to its mileage reduction, but MBUF raises total cost by about 5 percent even though driving falls, because the per-mile fee is collected through the tolls. This is the intended effect of the policy, where drivers pay more per mile and respond by driving somewhat less.

To isolate how much the land-use response contributes to these results, we recompute each policy effect on the BASE land-use evolution, applying only the travel policy while holding land use at the BASE trajectory, and compare it with the fully co-simulated effect in which both land use and transportation follow the scenario. Table~\ref{tab:cosim-effect} reports the 2045 VMT and VHT effects both ways for the seven-county total and for each county. The co-simulated column is the effect with the coupled land-use response, the uncoupled column holds land use on the BASE trajectory, and the contribution column is the co-simulated percent change minus the uncoupled percent change, which isolates the land-use response. On the Overall the two are close, within half a percentage point on every metric, so the land-use feedback is a small correction on the regional travel totals over this horizon. The Overall understates the feedback, however, because county movements of opposite sign cancel when summed. At the county level the VHT contribution reaches $1.4$ points under TELE in Cook, where households leaving the thinning core deepen the reduction, and $1.9$ points in Lake in the opposite direction. Under MBUF it is largest in the peripheral growth county of Kendall, where the mileage fee draws households toward the accessible core and deepens the VMT and VHT reductions by $1.8$ and $1.4$ points, while nearby Kane moves the other way. Coupling is therefore a relatively minor correction for regional accounting but material at the sub-regional scale. By the same logic, the contribution should grow at finer spatial scales, since the within-county movements that partly cancel in the county total would be resolved at the tract or block level.

\begin{table}[!ht]
\caption{Effect of each policy on 2045 VMT and VHT, as percent change from BASE.}
\label{tab:cosim-effect}
\centering
\tabfit
\begin{tabular}{llrrrrrr}
\hline
 & & \multicolumn{3}{c}{\textbf{VMT}} & \multicolumn{3}{c}{\textbf{VHT}} \\
\cline{3-5}\cline{6-8}
\textbf{Scenario} & \textbf{Area} & \textbf{Co-sim.} & \textbf{Uncoupled} & $\Delta$ & \textbf{Co-sim.} & \textbf{Uncoupled} & $\Delta$ \\
\hline
TELE & \textbf{Overall} & $-12.5$ & $-12.4$ & $-0.1$ & $-17.5$ & $-17.0$ & $-0.5$ \\
\cline{2-8}
      & Cook    & $-8.3$  & $-7.8$  & $-0.5$ & $-12.7$ & $-11.3$ & $-1.4$ \\
     & DuPage  & $-14.1$ & $-14.5$ & $+0.4$ & $-18.4$ & $-18.5$ & $+0.1$ \\
     & Kane    & $-15.1$ & $-15.4$ & $+0.3$ & $-22.0$ & $-22.5$ & $+0.6$ \\
     & Kendall & $-19.5$ & $-18.9$ & $-0.6$ & $-25.5$ & $-25.1$ & $-0.4$ \\
     & Lake    & $-17.0$ & $-18.3$ & $+1.2$ & $-27.3$ & $-29.2$ & $+1.9$ \\
     & McHenry & $-21.5$ & $-22.1$ & $+0.6$ & $-29.1$ & $-29.6$ & $+0.5$ \\
     & Will    & $-16.6$ & $-16.6$ & $-0.1$ & $-20.7$ & $-21.0$ & $+0.3$ \\
\hline
MBUF & \textbf{Overall} & $-2.9$  & $-2.7$  & $-0.3$ & $-3.1$  & $-3.0$  & $-0.1$ \\
\cline{2-8}
      & Cook    & $-3.0$  & $-2.5$  & $-0.6$ & $-3.6$  & $-3.1$  & $-0.5$ \\
     & DuPage  & $-2.8$  & $-3.0$  & $+0.2$ & $-2.3$  & $-2.9$  & $+0.6$ \\
     & Kane    & $-2.9$  & $-3.5$  & $+0.5$ & $-3.5$  & $-4.7$  & $+1.2$ \\
     & Kendall & $-5.5$  & $-3.7$  & $-1.8$ & $-5.9$  & $-4.5$  & $-1.4$ \\
     & Lake    & $-1.9$  & $-2.6$  & $+0.7$ & $-2.8$  & $-2.6$  & $-0.1$ \\
     & McHenry & $-2.8$  & $-2.9$  & $+0.2$ & $-1.5$  & $-2.0$  & $+0.5$ \\
     & Will    & $-2.9$  & $-2.2$  & $-0.6$ & $-1.2$  & $-1.5$  & $+0.3$ \\
\hline
\end{tabular}
\end{table}

\subsection{Building energy response}\label{subsec:results-buildingenergy}

In this subsection, we first characterize the daily occupancy the building stock produces, followed by the validation of the simulated energy intensity against measured benchmarking data. We then compare the annual energy of the two policy scenarios against BASE, and analyze where the difference comes from at the end.

\begin{figure}[!ht]
  \centering
  \includegraphics[width=\figw{0.75}]{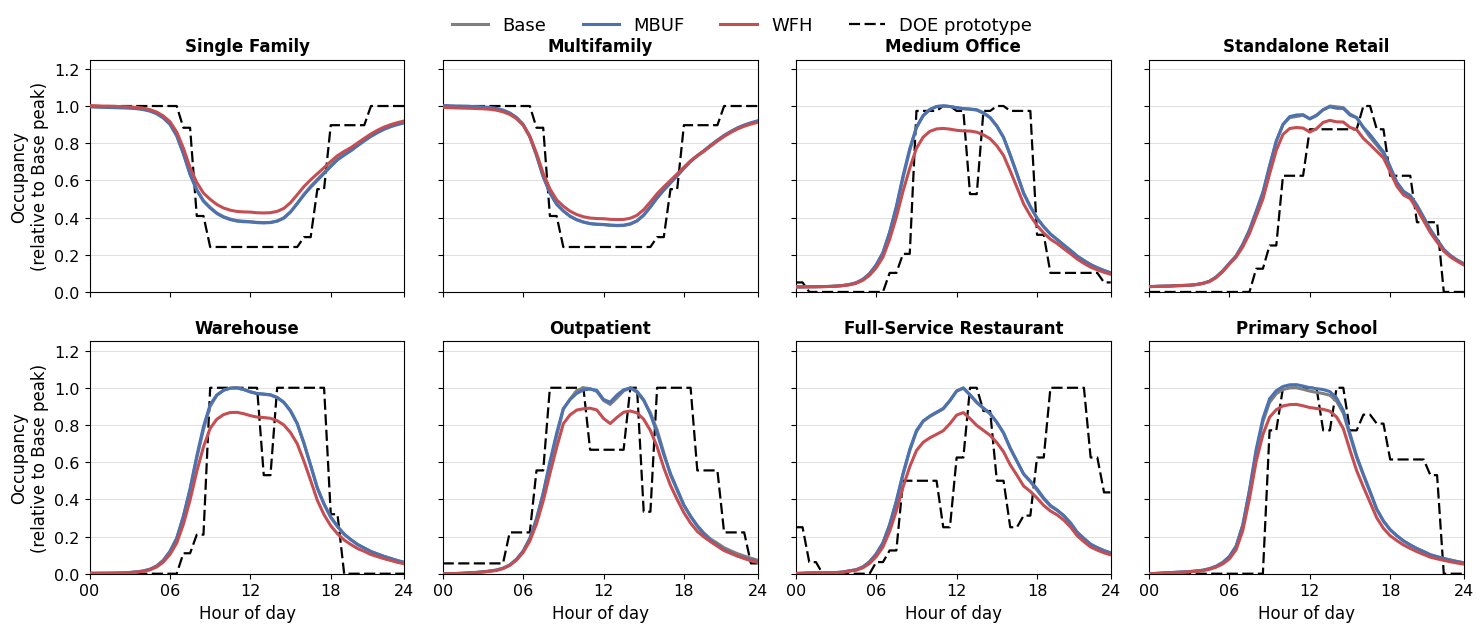}
  \caption{Normalized daily occupancy schedules for eight building types in Chicago 2045 under three scenarios, plotted against the DOE prototype reference.}
  \label{fig:occupancy}
\end{figure}

Figure~\ref{fig:occupancy} shows the occupancy schedules for eight prototypes in Chicago 2045 under three scenarios. Across all types, the transportation-informed schedules broadly follow the DOE prototype's occupancy schedule \cite{DOE_PrototypeModels}. Telecommuting is the only scenario whose schedules depart visibly from BASE, raising midday occupancy in single-family and multifamily homes while lowering it in the other building types.

\begin{figure}[!ht]
  \centering
  \includegraphics[width=\figw{0.8}]{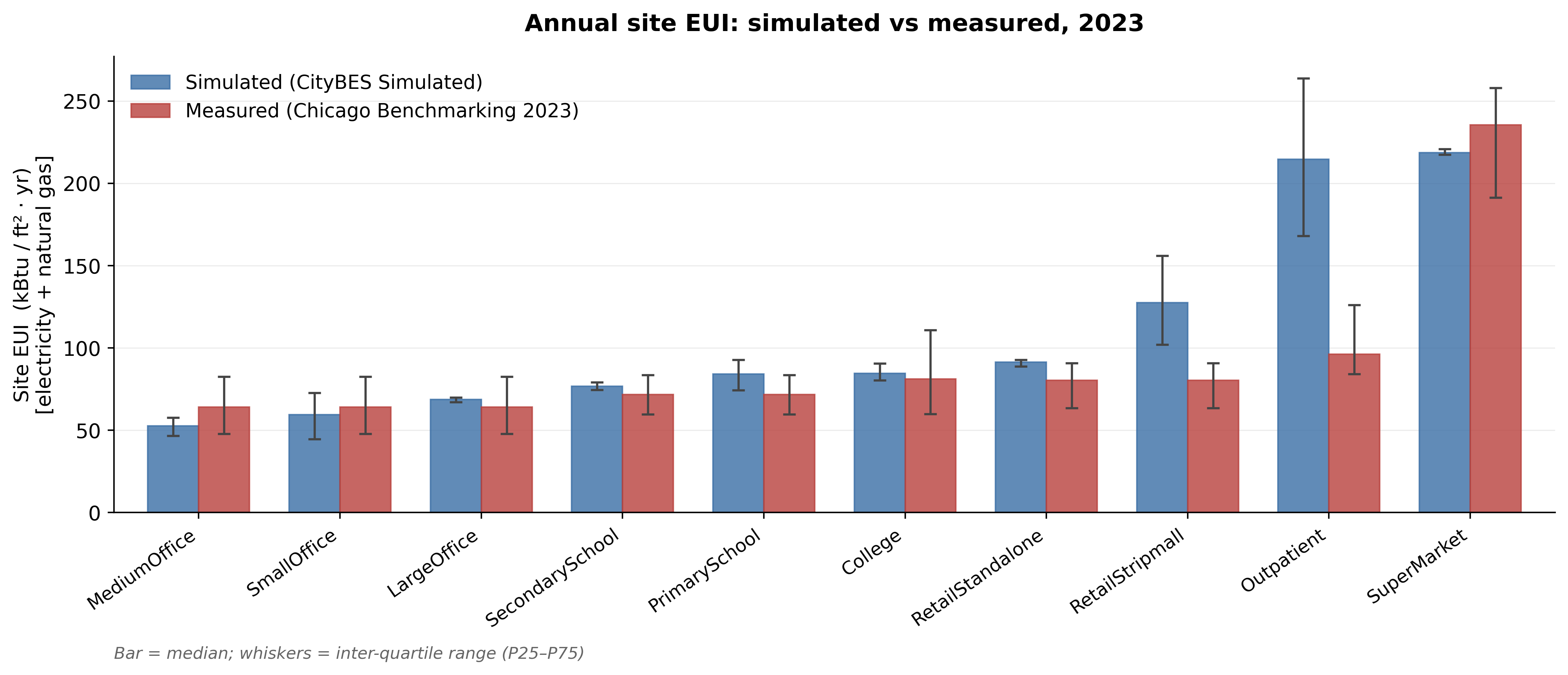}
  \caption{CityBES simulation versus City of Chicago Energy Benchmarking 2023.}
  \label{fig:eui}
\end{figure}

Figure~\ref{fig:eui} compares the simulated annual site energy use intensity (EUI) against the measured City of Chicago Energy Benchmarking data \cite{Chicago_Benchmarking_2023}. For most building types the simulated median falls within the measured inter-quartile range. However, the agreement is weaker for a few building types, most visibly Outpatient and Retail Strip Mall, which are overestimated and flagged for the targeted recalibration discussed in Section~\ref{sec:discussion}.  

\begin{figure}[htbp]
  \centering
  \includegraphics[width=\figw{0.8}]{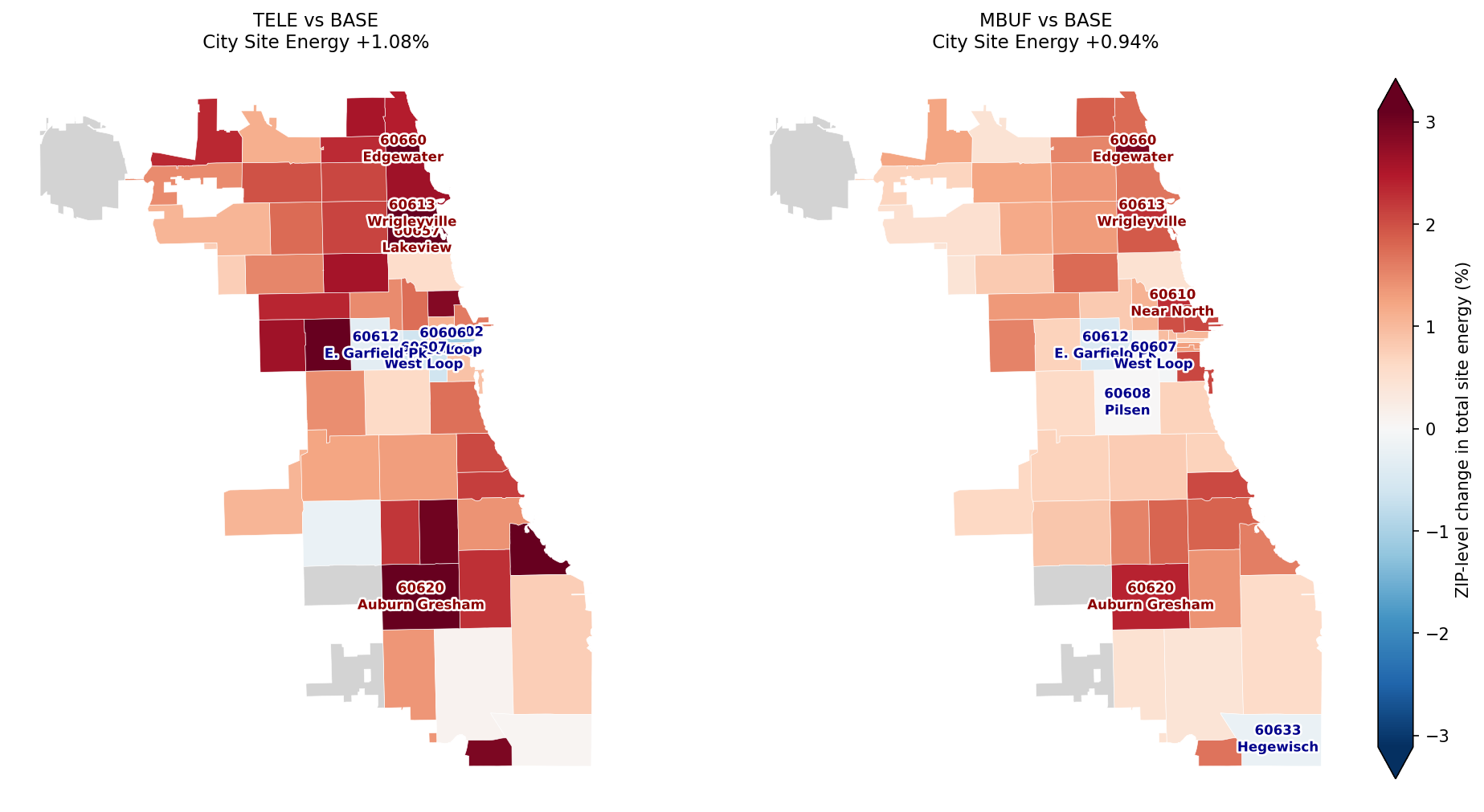}
   \caption{Zip-level percent change in total site energy usage relative to BASE, 2045.}
  \label{fig:zip-diff}
\end{figure}

\begin{figure}[htbp]
  \centering
  \includegraphics[width=\figw{0.8}]{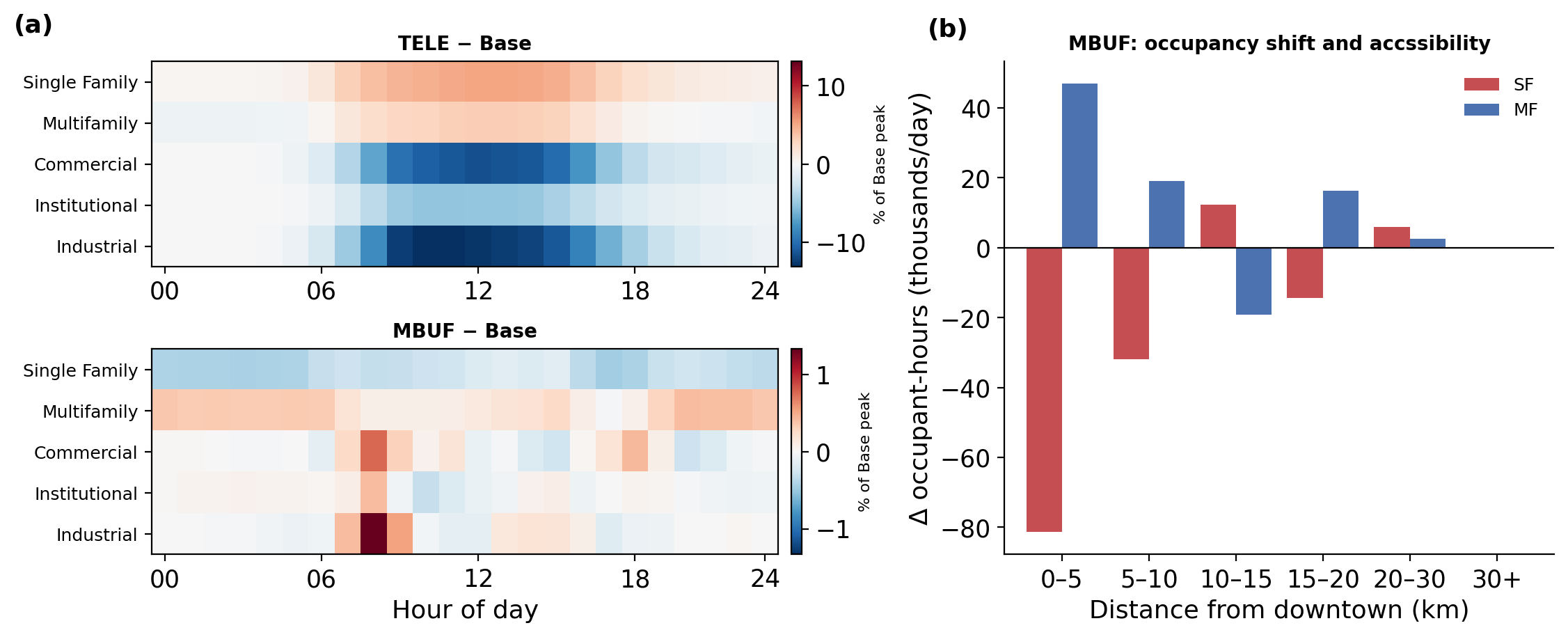}
  \caption{(a) hourly occupancy change by sector; (b) change in daily occupant-hours from BASE under MBUF, single-family versus multifamily, by distance from downtown}
  \label{fig:occ_combo}
\end{figure}

We now turn to the annual energy each scenario produces. In 2045, both policies raise citywide site energy relative to BASE, by 1.08\% under TELE and 0.94\% under MBUF. This partly reflects the lower VHT under both policies, because an individual who is not traveling is assumed to be inside a building, so reduced travel translates into more time spent indoors. Figure~\ref{fig:zip-diff} shows the rise is spread unevenly, with the residential North Side and lakefront gaining under TELE while the downtown area barely moves. The occupancy shift in Figure~\ref{fig:occ_combo}(a) is consistent with this pattern, with TELE moving occupants out of the workplace sectors and into single-family and multifamily homes. MBUF raises energy as well, though by less and with a more subdued occupancy signal, as the lower panel of Figure~\ref{fig:occ_combo}(a) shows occupants spending slightly more time in every sector except single-family. Figure~\ref{fig:occ_combo}(b) points to an accessibility mechanism, with occupants concentrating into the central multifamily stock within five kilometers of downtown. Multifamily is more than 20\% more energy-intensive per floor area than single-family at equal vintage in the prototype buildings, so the added occupancy loads more energy onto multifamily than it removes from the single-family homes people leave, and citywide energy rises.

\section{Discussion}\label{sec:discussion}

The platform reproduces a coherent, expected-sign response to both policies without any behavior being imposed directly. Because the scenarios reach land use only through the POLARIS skims and reach the travel model only through the policy inputs, the decentralization under telecommuting and the recentralization under the mileage fee are emergent, not assumed. The magnitudes are modest over the 2025--2045 horizon, which is reasonable given that anticipated regional growth is low and fixed, the fee is a moderate six cents per mile, and the horizon is roughly two decades. This effect is partly structural: households and jobs respond to altered accessibility only when they would otherwise relocate, so in a given year, only a minority of the population is exposed to the policy's effects. In reality, accessibility changes would also affect relocation rates and could induce workplace change without a corresponding residential move, channels that the platform does not yet represent. The value of the result therefore lies in the direction and spatial coherence of the response rather than its magnitude, and the modest scale of effects should be read as a conservative estimate rather than a null result.

A central question for a platform of this kind is whether coupling land use to transportation changes the answers a transportation-only study would give. Our comparison of the co-simulated runs against the uncoupled runs, in which the policy is applied on the BASE land-use evolution, shows that coupling is consequential, and most so at the county level. For the aggregate travel totals over this horizon the land-use feedback is a small correction, at most half a percentage point on any metric, so a transportation-only model would recover most of the regional VMT and vehicle-hour impact of either policy. At the regional level the land-use feedback reinforces rather than offsets the direct policy effect. Under each policy the resulting reallocation of households relative to the accessible core shifts travel in the same direction as the policy itself, so the coupled reduction in VMT and vehicle-hours is modestly larger than the transportation-only reduction. The correction is larger at the county level, where the offsetting movements no longer cancel, so coupling matters most for sub-regional analysis and for pricing policies that change accessibility, and least for near-term regional VMT accounting.

Two further caveats bound these findings. Future-year network speeds are elevated and are flagged for recalibration, which may affect the level of the travel metrics. The regional control totals are also held fixed across scenarios, so the analysis captures spatial redistribution of activity but not scenario-driven changes in aggregate regional growth.

The platform's components are general in design. UrbanSim's block-level model is built from Census and employment data available for many regions, POLARIS has been deployed in multiple contexts, and CityBES's prototype-and-vintage approach follows national reference building types. The workflow is therefore transferable, and the binding constraints are data and calibration rather than the approach. The building-to-activity map and the density thresholds in the building generation module were derived from the Chicago building stock, so a new region needs an equivalent footprint-level building dataset to rebuild them. The building energy validation relied on Chicago's energy benchmarking ordinance, and regions without disclosure requirements would need an alternate data source. The land-use model's calibration requires observed block-level distributions at two points in time, which the decennial Census supplies nationally, though the calibration window is limited accordingly and the model benefits from local land-use policy inputs and fine-tuning.

\section{Conclusion}\label{sec:conclusion}
The main contribution of this paper is the successful development of a co-simulation platform that couples land use, transportation, and building energy for the Chicago region on a shared geospatial data layer, the first to integrate all three sectors simultaneously. We demonstrated that the platform runs end to end and produces internally consistent multi-sector forecasts. Furthermore, we showcased its capabilities through two policy applications, a high-telecommuting scenario and a mileage-based user fee, benchmarked against a business-as-usual case through 2045. 

As detailed in the Discussion, the coupled system produced coherent, expected-sign responses to both policies, telecommuting decentralizing activity and the mileage fee recentralizing it, and the land-use feedback on regional travel, while small in aggregate, grew at finer spatial scales to become essential for analysis at the county level or below. On the building side, carrying transportation-informed occupancy and the evolving building stock into a physics-based energy model exposed a coherent reallocation of building energy under both policies, one that fixed-stock or single-sector modeling would miss.

Future work is planned across the three components. On the transportation side, POLARIS will be recalibrated to correct the elevated future-year network speeds; on the land use side, UrbanSim will be extended to represent telecommuting-driven residential choices beyond travel-time effects. On the building side, the prototypes flagged by the benchmarking comparison as biased for certain building types, notably outpatient facilities and retail strip malls, will be refined. With these improvements, the co-simulation platform will support more detailed studies of automation and teleworking, further demonstrating its broad benefits and capabilities.

%% file: backmatter.tex
\section*{Acknowledgments}
This report and the work described were sponsored by the U.S. Department of Energy (DOE) Transportation Technologies Office (TTO) under the Integrated Transportation and Energy Cross-Sectoral System of Systems at Scale (ITES4) project. Melissa Rossi, DOE Office of Critical Minerals and Energy Innovation (CMEI) managers, played an important role in establishing the project concept, advancing implementation, and providing guidance. The authors would also like to thank the Chicago Metropolitan Agency for Planning for providing access to a version of its UrbanSim model and associated input data. The model was modified for this analysis; the results and interpretations presented here do not represent official Agency forecasts or views. The submitted manuscript has been created by the UChicago Argonne, LLC, Operator of Argonne National Laboratory (Argonne). Argonne, a U.S. Department of Energy Office of Science laboratory, is operated under Contract No. DE-AC02-06CH11357. The U.S. Government retains for itself, and others acting on its behalf, a paid-up nonexclusive, irrevocable worldwide license in said article to reproduce, prepare derivative works, distribute copies to the public, and perform publicly and display publicly, by or on behalf of the Government.


\section*{Author Contributions}
The authors confirm contribution to the paper as follows: study concept and design: all authors; data preparation: G. Nair, Y. Jiang, S. Maurer, J. Cook; methodology: all authors; software development: G. Nair, Y. Jiang, S. Maurer, J. Cook, J. Auld, T. Hong, A. Besharati, P. Waddell; analysis and interpretation of results: all authors; manuscript preparation: G. Nair, Y. Jiang, S. Maurer, N. Khan, T. Hong, A. Besharati, P. Waddell. All authors reviewed the results and approved the final version of the manuscript.

\section*{AI Use Disclosure}
AI tools developed by Anthropic and OpenAI were used to assist with software development, data analysis, and refinement of the initial manuscript draft, including improvements to language, clarity, concision, and grammar. All code and editorial revisions generated with the assistance of these AI tools were carefully reviewed by the authors. The authors take full responsibility for the final content of this publication.

\section*{Declaration of Conflicting Interests}
The authors declared no potential conflicts of interest with respect to the research, authorship, and/or publication of this article.

\section*{Funding}
This research was supported by the U.S. Department of Energy (DOE) Transportation Technologies Office (TTO) under the Integrated Transportation and Energy Cross-Sectoral System of Systems at Scale (ITES4) project.